\documentclass[%
aip,
amsmath,amssymb,
reprint,%
]{revtex4-2}

\usepackage{graphicx}
\usepackage{dcolumn}
\usepackage{bm}

\usepackage[utf8]{inputenc}
\usepackage[T1]{fontenc}
\usepackage{mathptmx}
\usepackage{etoolbox}
\usepackage[mathscr]{euscript}
\newcommand*{\rom}[1]{\expandafter\@slowromancap\romannumeral #1@}

\usepackage {xcolor}
\usepackage{chemformula}

\def\bef{\begin{framed}}
	\def\eef{\end{framed}}
\def\be{\begin{equation}}
	\def\ee{\end{equation}}
\def\ber{\begin{eqnarray}}
	\def\eer{\end{eqnarray}}

\def\nn{\nonumber}

\makeatletter
\def\@email#1#2{%
	\endgroup
	\patchcmd{\titleblock@produce}
	{\frontmatter@RRAPformat}
	{\frontmatter@RRAPformat{\produce@RRAP{*#1\href{mailto:#2}{#2}}}\frontmatter@RRAPformat}
	{}{}
}%
\makeatother
\begin{document}
	
	\preprint{AIP/123-QED}
	
	\title{Thermal magnetoresistance from magnon scattering from a domain wall in an antiferromagnetic insulator}
	\author{Ehsan Faridi}
	\email{efbmm@mail.missouri.edu}	 
	\affiliation {Department of Physics and Astronomy, University of Missouri, Columbia, Missouri 65211, USA}
	\author{Se Kwon Kim}%
	\affiliation{ 
	 Department of Physics, Korea Advanced Institute of Science and Technology, Daejeon 34141, Korea
	}%
	
	\author{Giovanni Vignale}
	\email{vignaleg@missouri.com}
	\affiliation{ The Institute for Functional Intelligent Materials (I-FIM), National University of Singapore, 4 Science Drive 2, Singapore 117544
	 }%
 \affiliation {Department of Physics and Astronomy, University of Missouri, Columbia, Missouri 65211, USA}
	
	\date{\today}
	
	\begin{abstract}
		We theoretically investigate magnon heat transport in an  antiferromagnetic (AFM) insulator containing a  domain wall (DW) in the presence of a magnetic field  applied along the easy axis. We show that  the intrinsic  spin of the DW couples to the external magnetic field which modifies the transmission of spin wave (SW) through the DW. Applying the magnetic field    lifts the degeneracy between two AFM magnon modes and results in different occupation numbers for the two magnon modes. Combined with the finite reflection of a narrow domain wall, this is found to have a significant impact on the magnon heat transport, giving rise to thermal magnetoresistance. Our findings suggest that an  AFM DW can be used as a controllable element for regulating the magnon heat current in magnonic devices through the application of a magnetic field.

	\end{abstract}
	
	\maketitle

	Thermal conductivity, a key parameter for describing heat transport in solids, is generally governed by electrons and phonons  in metals and phonons in insulators \cite{pekola2021colloquium}. In magnetically ordered solids, energy can also
	be carried by magnons, the quanta of the collective oscillations of the spin. \cite{vilela2022thermally,polash2020magnon,han2024magnonics}. 	 
    Magnon contributions to heat transport
	have been demonstrated in both insulators \cite{rives1969effect,wang2022magnon,cui2024anisotropic,tang2024magnon,prasai2017ballistic}  and metals\cite{natale2021field,gronert1988evidence}. For instance, it has been found that  magnons contribute more than half of the heat current in ferrimagnetic YIG  
 \cite{rives1969effect}.  In addition, the time-domain thermoreflectance measurements reveal that the magnons significantly contribute to the heat conduction even in ferromagnetic metals at room temperature \cite{hirai2024nonequilibrium}. 

Magnon mediated heat transport  in  ferromagnets has been a topic of interest for decades \cite{bauer2012spin,douglass1963heat}.  Magnons interact directly with magnetic fields, resulting in a red or blue shift in their dispersion \cite{shen2020driving,rezende2019introduction}. This shift alters the thermal occupation of magnon states, enabling control over the material's thermal conductivity through the Zeeman effect.   Furthermore, it is known that scattering of magnons by ferromagnetic DW, connecting regions in which the order parameter has opposite signs, is a major mechanism of thermal resistance, reducing the thermal current\cite{yan2012magnonic}. 

However, a similar study has not been done for AFM insulators, where magnons exist in two states of polarizations which respond differently to a magnetic field. Interest in AFM materials has grown in recent years, driven by advances in AFM spintronics \cite{jungwirth2016antiferromagnetic,baltz2018antiferromagnetic}.   In this paper, we demonstrate that the external magnetic field couples to an AFM  DW, modifying its net spin and influencing the magnon transmission coefficients of both polarizations. This effect, combined with the field-induced changes in the magnon dispersion relation, allows for control of magnon heat transport via an applied magnetic field -- an effect that we describe as thermal magnetoresistance from magnon scattering from domain walls.


	As a model system, we consider
 a 1D nanowire which contains a domain wall between two semi-infinite homogeneous spin chains with opposite orientations of the Neel order parameter. The interaction between spins can be described by the Hamiltonian consisting of  exchange, magnetic anisotropy and Zeeman energies in the form
	\ber
	\mathcal{H}&=&J\sum_{n}\textbf{S}_n \cdot \textbf{S}_{n+1}-\textit{D}\sum_{n}(S_n^z)^2-\gamma\hbar \sum_{n}\textbf{B} \cdot \textbf{S}_n   
	\eer
	where  $S_ n$ is the  spin on the lattice site $n$, $\textbf{B}$
	is the applied magnetic field , $J$ is the exchange constant of the
	interaction between spins $\textbf{S}_n$ and $\textbf{S}_{n+1}$  and D is the uniaxial anisotropy constant. We assume $J>0$ so that the exchange interaction favors antiparallel oreientation of neighboring spins, leading to antiferromagnetism. 
	In the presence of a magnetic field less than spin-flop field i.e. $\gamma\hbar B < \gamma\hbar  B_{sf} = 2\sqrt{D(D+2J)}$,  below which the AFM phase is stable, the equilibrium configuration of a DW can be obtained by minimizing the Hamiltonian with respect to the polar angles $\theta_n$ describing the equilibrium orientation of $\textbf{S}_n$. Requiring $\delta \mathcal{H}/\delta\theta_n=0$ gives  the equilibrium configurations for the DW:
	\be
	\sin(\theta_n-\theta_{n+1})+\sin(\theta_n-\theta_{n-1})-\frac{D}{J}\sin2\theta_n-\frac{B\gamma\hbar}{J}\sin\theta_n=0.
	\ee

	A numerical solution of Eq. (2) is shown in Fig. 1(a) for $D/J=0.1$ and $\gamma B/J=0.3$.  In the calculation we consider a DW with even pairs of spins  which have lower energies compared to those with odd pairs of spins \cite{hilzinger1972spin, faridi2024control, faridi2022atomic}. We assume that on left side of the DW,  $n < 1$,  the spins of sublattice A(B) which are located at odd (even) sites are pointing in the $z$ ($-z$) direction and on the right side of the DW, $n > 2N$,  the spins of sublattice A(B)  are pointing in the $-z$ ($z$). In the DW region, $1\leq n\leq 2N$,  each of the $N$ ``unit cells" composed of sites $(1,2)$, $(3,4)$...$(2N-1,2N)$,  has a net spin even in the absence of a magnetic field \cite{tveten2016intrinsic}. The total spin within each unit cell increases near the center of the DW and rapidly decreases as we approach the homogeneous region where the homogeneous alignment of spins is restored. As the magnetic field is switched on the spins tend to be aligned in the direction of the field which results in increasing  of the total spin inside the DW (see Fig. 1 (b)).   
	
 	\begin{figure}
 	\centering
 	\includegraphics[width=86mm]{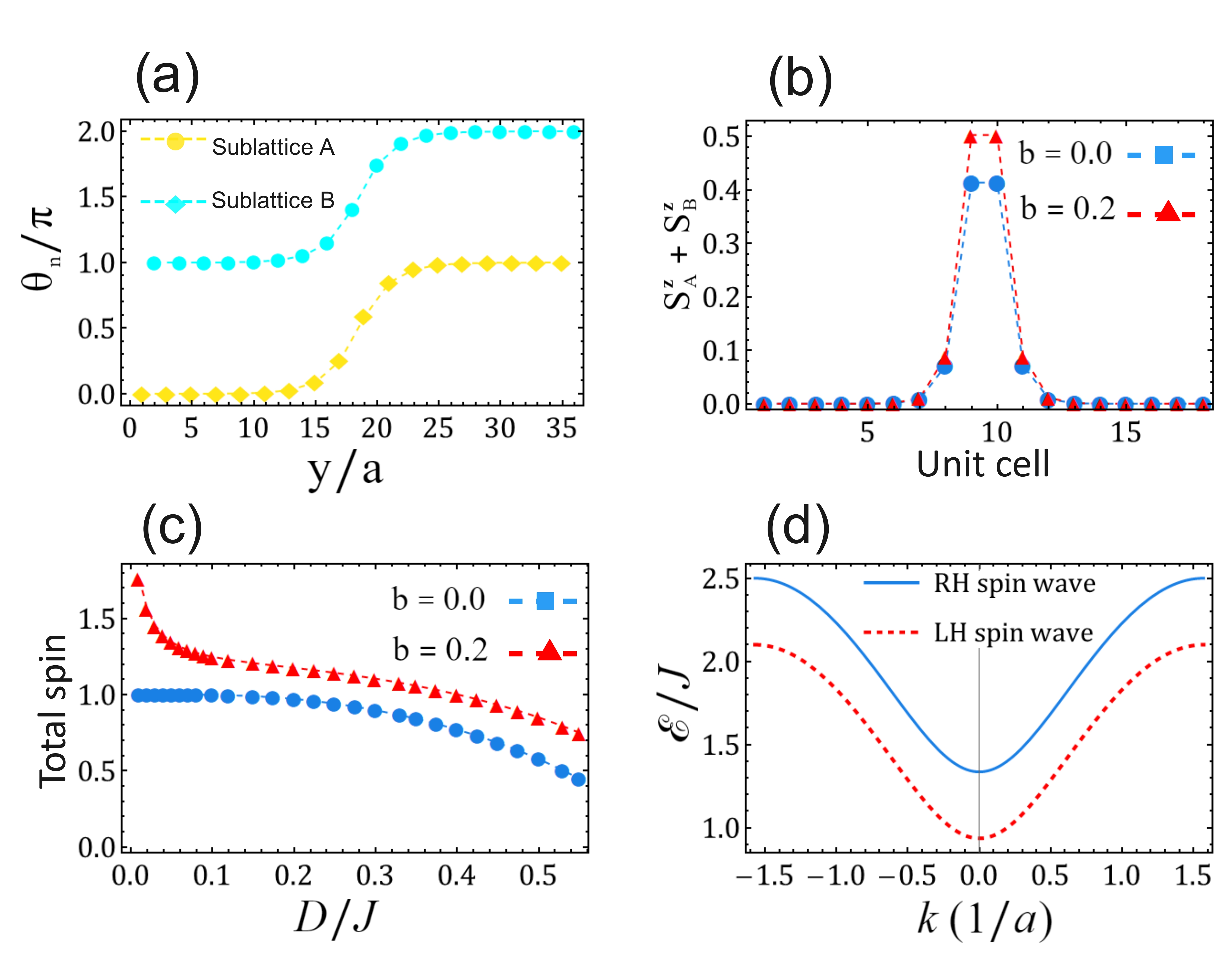}\caption { (a) Exact DW profile with $ d\equiv D/J = 0.15$ and $b\equiv\gamma B\hbar/J=0.2$. The blue square  represent the angles of the spins at sublattice A, and the yellow discs represent those at sublattice B. (b) Net spin in a AFM DW at each unit cell. (c) Total spin in a AFM DW as a function of $D/J$ in the absence of a field (blue dots) and with an applied field of $\gamma B_0\hbar/J=0.2$. (d) Dispersion of a RH and a LH SWs as a function of wave vector with $b = 0.2$ and $d = 0.15$.  }	
 \end{figure} 
	
	We shall study the dynamics of spins in a local coordinate in which the $z$ axis is chosen to point along the spin equilibrium direction. The dynamic of spins is expressed by equation of motion:
	
	\begin{equation}
		\hbar\dot{S}_{n, \alpha}=i [\mathcal{H},S_{n\alpha}]\,,
	\end{equation}
	Here $\alpha$ and $\beta$ take values in  $\{-,+\}$ and $S_{n,+}$ and $S_{n,-}$ are the chiral components of the spin
	deviation. Thus the linearized equation of motion reads:
	\begin{equation} \label{schero_spin_wave}
		\hbar\omega S_{n\alpha} = \sum_{n'\alpha\beta}H_{n\alpha,n'\beta} S_{n'\beta}\,.
	\end{equation}
	Here the diagonal part of the SW Hamiltonian is expressed as:
	\begin{eqnarray}
		H_{n\alpha,n \beta}&=& -JS(c_{n-1}  +c_n)[\sigma_z]_{\alpha\beta} \nonumber \\  &+&   DS(2\cos^2 \theta_n-\sin^2\theta_n) [\sigma_z]_{\alpha\beta}\nonumber\\
		&+& DS\sin^2\theta_n[i\sigma_y]_{\alpha\beta}+\gamma\hbar B_0\cos\theta_n [\sigma_z]_{\alpha\beta}\,,
	\end{eqnarray}
	and the off-diagonal part is
	
	\begin{eqnarray}
		H_{n\alpha,n+1 \beta}&=&JS\left\{\frac{1+c_n}{2} [\sigma_z]_{\alpha\beta}+\frac{1-c_n}{2}[i\sigma_y]_{\alpha\beta}\right\}\,,\nonumber\\
		H_{n\alpha,n-1 \beta}&=&JS\left\{\frac{1+c_{n-1}}{2} [\sigma_z]_{\alpha\beta}+\frac{1-c_{n-1}}{2}[i\sigma_y]_{\alpha\beta}\right\}\,. \nonumber
	\end{eqnarray}
	
	where $c_n \equiv \cos(\theta_n - \theta_{n+1})$ and $\sigma_i$ are  Pauli matrices which act on spin deviations.
    
    To obtain eigenvalues and eigenvectors of Eq. (3) for the {\it homogeneous} AFM phase, we set  $\theta_n=0$ for odd $n$ and $\theta_n=\pi$  for even $n$. We find two orthogonal solutions.  The right-handed (RH) solution (written in the chiral basis) has eigenvalues  
	\be
	\mathcal{E}_{RH}=\gamma\hbar B+\hbar\omega_k ,
	\ee
	with eigenvectors
	\be
	u_{k\uparrow}^{(RH)}=N_{k,1}\begin{pmatrix} 
		1\\0  
	\end{pmatrix},
	~~ u_{k\downarrow}^{(RH)}= -N_{k,2}\begin{pmatrix}	0\\1  
	\end{pmatrix}\,,
	\ee 
    where   
    \be
    \hbar\omega_k= 2 \sqrt{D(2J+D)+J^2\sin^2 ka}
    \ee
    is the eigenvalue of the two degenerate modes in the absence of the magnetic field, $\uparrow (\downarrow)$ refer to direction of the spins at the sublattice A(B) and $N_{k,1(2)} = \sqrt{\frac{4(J+D)\pm 2\hbar\omega_k}{\hbar\omega_k}} $    are the amplitudes of the oscillation on the sites with  $\uparrow(\downarrow)$ spins.  
	Similarly, the left-handed (LH) solution has eigenvalues
	\be
	\mathcal{E}_{LH}=-\gamma\hbar B+\hbar\omega_k ,
	\ee
	with eigenvectors
	\be
	u_{k\uparrow}^{(LH)}=N_{k,2}\begin{pmatrix} 
		0\\1  
	\end{pmatrix},
	~ u_{k\downarrow}^{(LH)}=-N_{k,1} \begin{pmatrix}	1\\0  
	\end{pmatrix}\,.
	\ee 
	
	The LH-mode carries a pure spin in the $+z$ direction (parallel to the magnetic field) . That is why it has a smaller energy than the RH-mode which carries a pure spin in the $-z$ direction (anti-parallel to the magnetic field). 

\begin{figure}
\centering
\includegraphics[width=86mm]{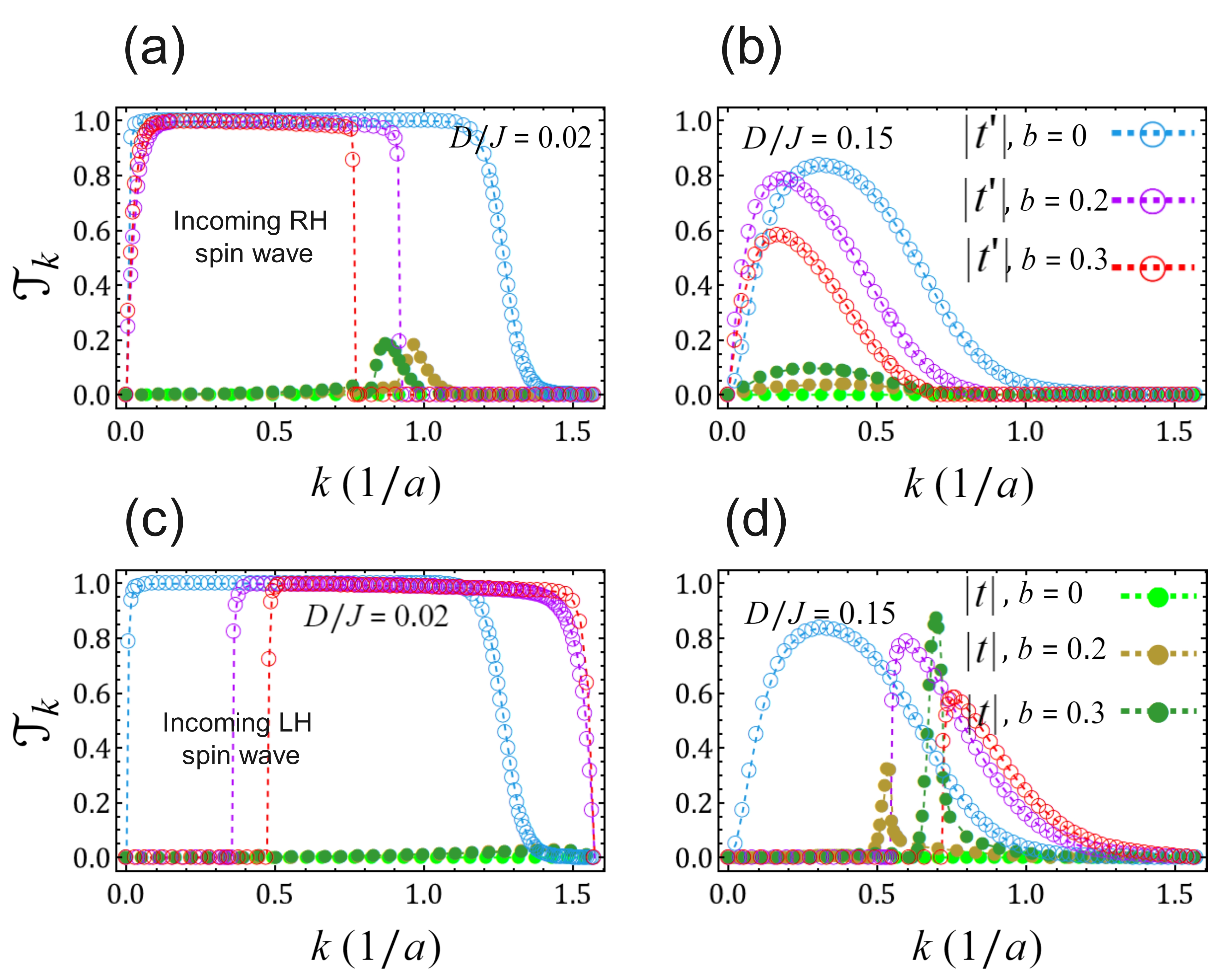}\caption{Transmission coefficients ($\mathscr{T}_k$) of an incoming RH  wave for various magnetic field values with parameters (a) $D/J=0.02$ , (b) $D/J=0.15$.  Filled circles indicate the transmission probability for a SW with the same polarization as the incoming wave, i.e. $|t_k|^2$ and empty circles represent the transmission probability for a SW with the opposite polarization i.e. $|t'_k|^2$. Panels (c), (d)  are like (a) and (b) respectively, but for an incoming LH wave.  Note that the transmission coefficients of the wave with the same polarization as the incoming wave (light green filled circles) is nearly zero when the field is absent.}	
			\end{figure}

\begin{figure*}
\centering
\includegraphics[width=180mm]{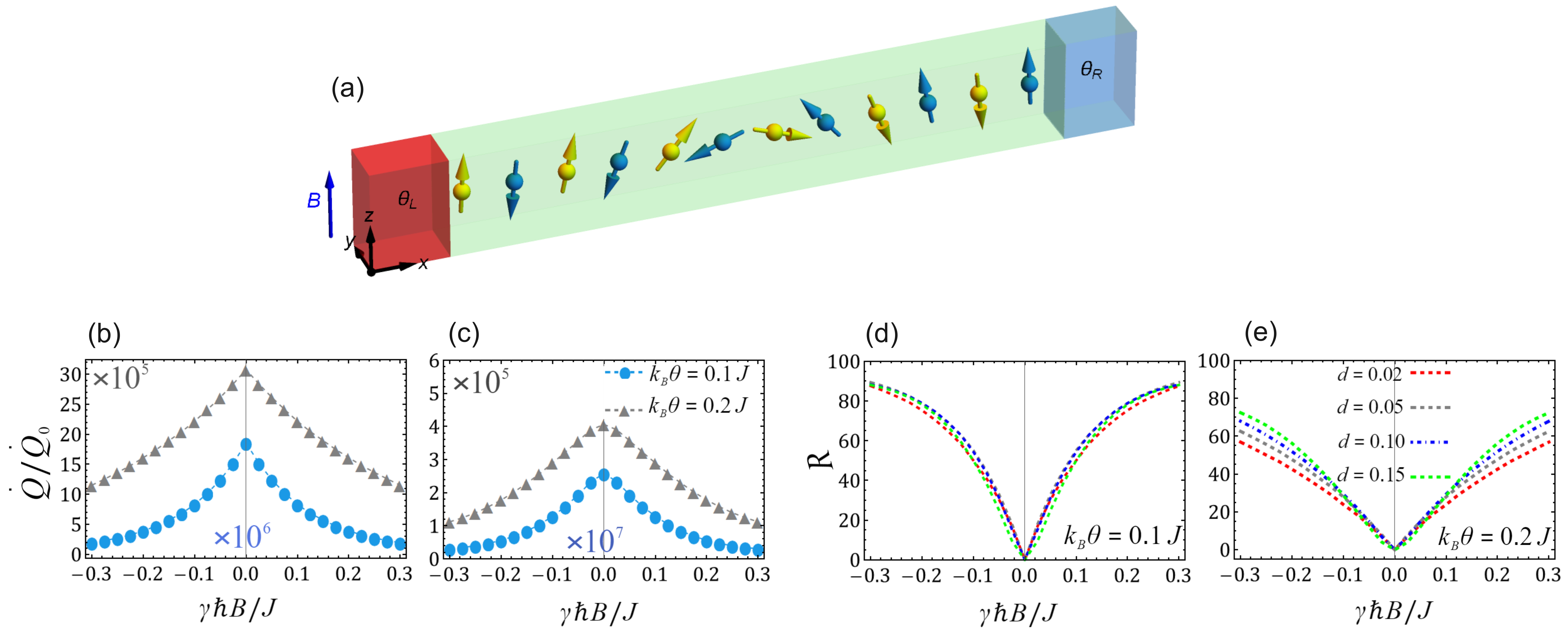}\caption{ (a) Schematic of an AFM nanowire containing a DW located between two heat sources. (b) Magnon-heat current across the DW  vs magnetic field for $D/J = 0.02$ at two different temperatures,  $k_B\theta = 0.1J$ (filled blue circles) and  $k_B\theta = 0.2J$ (filled gray triangles)  (c) Same as (b) for $D/J = 0.15$.  Here $\dot{Q}_0 = J^2/h $.  (d) Thermal magnetoresistance across the DW vs magnetic field for various values of $D/J$ at   $k_B\theta = 0.1J$  (e) same as (d) for $k_B\theta = 0.2J$.   The temperature difference between two sources is set at $k_B\Delta\theta=0.01 J.$  
  }\label{Fig3}
\end{figure*}

In order to calculate the scattering of SW upon a DW we employ the Green's function technique introduced in Ref \cite{faridi2022atomic}. According to Fig. 2 in the absence of the magnetic field and for a wide DW,  i.e. $D/J\ll1$, both RH and LH SWs pass through the DW with negligible reflection and appear on the right side of the DW with a reversed polarization. As the magnetic field switched on, the RH and LH SW show different behaviors upon the transmission through the DW. Note that for a particular polarization of an incoming SW, the transmitted SW is a superposition of both a RH and a LH SW. Let us first focus on an incoming RH SW.  As shown in Fig. 2(a,b), an incoming RH SW results in a transmitted SWs with two components. The first component, indicated by filled circles, maintains the same RH polarization as the incoming wave. The second component, indicated by empty circles, has the reversed LH polarization. As the applied field increases for a fixed value of anisotropy, the transmission with the same polarization rises due to increase in the net spin of each unit cell, while the transmission with the reversed polarization decreases. However, for larger values of anisotropy, the overall transmission probability decreases. In addition, 
for a wave vector larger than a certain value, i.e. $k>k_0^{(RH)}$ there is a cut off for transmission coefficients with the reversed polarization, beyond which the amplitude of SW with the reversed polarization is zero.
The reason is that, on the right side of the DW, when  $k>k_0^{(RH)}$, there is no magnon state     with the same wave vector and the same energy as the incoming wave. The value of $k_0^{(RH)}$ depends on the applied field and the anisotropy constant:

\be
k_0^{(RH)} = \sin^{-1}\left(\sqrt{(b - 1)^2 - 2b d}\right)
\ee
where $b\equiv\gamma\hbar B/J$ and $d\equiv D/J$.

Similar argument can be made for an incoming LH SW. As depicted in the second row of Fig. 2(c,d), when the magnetic field is increased, the transmission probability of the waves with the same polarization as the incoming wave increases, while the transmission probability for waves with the reversed polarization decreases. For larger values of $D/J$, in contrast to the incoming RH SW, an incoming LH SW exhibits larger transmission probability with  the same polarization. In addition, for the wave vector less than a certain value, i.e. $k<k_0^{LH}$, the SW with the reversed polarization experiences a complete reflection. For $k<k_0^{(LH)}$ the wave vector becomes purely imaginary and the transmitted wave takes the form of an evanescent wave. We can also obtain the expression for $k_0^{(LH)}$

\be
k_0^{(LH)} = \sin^{-1}\left(\sqrt{b^2 + 2 b \sqrt{d^2 + 2 d}}\right).
\ee

It is worth noting that for $D/J>2/3$ the DW becomes abruptly sharp which posses $U(1)$ symmetry, i.e. it is invariant under rotations in spin space around the $z-$axis. The symmetry persists even in the presence of an external field in the $z-$direction. Owning to this symmetry, the transmission coefficient of SW regardless of its polarization does not change by applying a magnetic field.

Reflection of SWs from a DW  modifies the magnon-heat current in a nano wire. We consider a nano-wire with length L which connects
two large heat reservoirs held at two temperatures $\theta_L$, $\theta_R$, with a constant difference $\Delta \theta = \theta_L-\theta_R > 0$. The temperature difference between two sides of the DW creates a magnon imbalance which leads to a magnon flow from left side to the right side of the DW. 
The heat current mediated by SW within the spin chain is described by Landauer-Buttiker formula\cite{meier2003magnetization,yan2012magnonic}:

			\ber\label{landauer}
			\dot{Q}_{DW}&=& \frac{ 1}{L}\int dk \rho_k   v_k   \mathcal{E}^{RH}_k\{ n_B(\mathcal{E}^{RH}_k,\theta_L)-n_B(\mathcal{E}^{RH}_k,\theta_R)\} T_k^{RH}
			\nn\\&+&\frac{ 1}{L}\int dk  \rho_k v_k    \mathcal{E}^{LH}_k\{ n_B(\mathcal{E}^{LH}_k,\theta_L)-n_B(\mathcal{E}^{LH}_k,\theta_R)\} T_k^{LH}.\nn\\
			\eer
Here $\rho_k=1/2\pi$ is the magnon density of states, $v_k=\partial\omega/\partial k=2aJ^2\sin( 2ka)/\hbar^2\omega_k$ is the group velocity of the SW, $T_k = |t_k|^2 + |t'_k|^2$ and $n_B(\omega_k,\theta _{L,R})=1/(e^{\omega/(k_b\theta_{L,R})}-1)$ is the Bose–Einstein distribution. The first (second) integral represent the contribution of an incoming RH (LH) SW in the heat current.  
			
In Fig. \ref{Fig3}(b,c), the SW mediated heat current  is shown as a function of an external field   for two values of $D/J$ and two sets of temperatures which are small enough to minimize the phonon contribution to the heat current. As we can see  the magnon heat current monotonically  decreases with increasing the applied field due to the reduction of transmitted SW.  For larger values of $D/J$, where the DW becomes narrower, the transmission probability  decreases which leads to a lower magnon heat current compared to the values observed at lower $D/J$. For a larger temperature i.e. $k_B\theta = 0.2J$,    the magnon heat current significantly increases due to the increasing of the magnon imbalance at two sides of the DW.  At higher temperatures, e.g., for $k_B\theta \rightarrow J$, the magnon density becomes very large and the simple linear SW theory breaks down due to magnon-magnon interaction.

Numerically, the  thermal magnetoresistance can be characterized by the value:
			\be
			 R=\frac{\dot{Q}(0)-\dot{Q}(B)}{\dot{Q}(0)}\,,
			\ee
which depends on the $D/J$ ratio and the magnetic field $B$. As we can see from Figs.~\ref{Fig3}(d,e), thermal magnetoresistance is larger at  low temperatures, e.g., $k_B\theta = 0.1J$. 
At these temperatures, we find that the magnetoresistance has negligible dependence on the $D/J$ ratio. However, for larger temperatures, we find that the magnetoresistance grows with increasing $d=D/J$ (i.e., for narrower DWs).  This can be attributed to a stronger dependence of the transmission coefficients on a magnetic field in spite of the fact that there is no change in the dispersion of the transmitted waves.  
  As a practical example, we consider two  AFMs, MnF$_2$ and FeF$_2$ with the N\'eel temperatures of $67.3$K and $78.4$K respectively \cite{shapira1976magnetostriction,crc_handbook_89}. The magnetic interactions in  MnF$_2$ and FeF$_2$ are well described by Eq. (1) with the similar  nearest neighbor exchange constant   $J\approx 0.152 (meV)$ and $J \approx 0.156(meV)$ respectively. In MnF$_2$, the  anisotropy constant is  small $D \approx 0.002$ (meV) due to the lack of orbital angular momentum in the Mn$^{+2}$ ions, resulting in a wider DW. On the other hand, FeF$_2$ exhibits a relatively large anisotropy constant of $D \approx 0.06$ (meV), due to the finite orbital angular momentum of Fe$^{2+}$ ions leading to a narrower DW\cite{okazaki1964neutron,niira1954magnetic,jaccarino1983temperature,barak1978magnetic}.  Using these parameters in the Eq. (\ref{landauer})  we can estimate the reduction in the magnon-heat current caused by the reflection of the SW from a DW in both MnF$_2$ and FeF$_2$  . Note that, we assume that in the absence of the  DW, all  magnons are able to reach to the cold source. In the case of MnF$_2$ where $D/J\approx 0.02$  there is no detectable reduction in the magnon heat current in  the presence of a DW. However, applying a magnetic field in the amount of $\gamma\hbar B = 0.05 J \approx 0.06(T)$ could reduce the heat current and gives  $R(B)=30\%$. In contrast, for FeF$_2$ where $D/J\approx0.3$, the DW is narrower, with a length approximately   $\sqrt{D_{\text{MnF}_2}/D_{\text{FeF}_2}}\approx 0.26$ times that of MnF$_2$. Here, the presence of the DW alone can reduce the heat current by a significant percentage. Applying the same magnetic field,  can further reduce the heat current by $R=40\%$.

\section{Summary}
			
In this paper, we have theoretically studied the effect of  magnetic field on AFM DW and the  heat current carried by magnons. We have demonstrated that presence of a DW along with the magnetic field could reduce magnon heat current significantly.    Our results suggest that AFM DWs can serve as tunable components for controlling magnon heat current in magnonic devices via magnetic fields, offering potential applications in the design of next-generation spintronic devices.

\begin{acknowledgments}
	S.K.K. was supported by Brain Pool Plus Program through the National Research Foundation of Korea funded by the Ministry of Science and ICT (NRF-2020H1D3A2A03099291), by the National Research Foundation of Korea (NRF) grant funded by the Korea government (MSIT) (NRF-2021R1C1C1006273), and by the National Research Foundation of Korea funded by the Korea Government via the SRC Center for Quantum Coherence in Condensed Matter (NRF-RS-2023-00207732).
	\end{acknowledgments}
\bigskip

    \bibliography{Draft_Magnon_Heat_Ref}
			
			
		\end{document}